\documentclass[conference]{IEEEtran}

\IEEEoverridecommandlockouts  

\usepackage[mathscr]{euscript}
{}
{}
\usepackage{nomencl}
\makenomenclature
\usepackage{hyperref}
\usepackage{amsmath}
\usepackage{enumitem}
\usepackage{graphicx,epstopdf}
\usepackage{multicol}
\usepackage{tabularx}
\usepackage{longtable}
\usepackage{textcomp}
\usepackage{multirow}
\usepackage{booktabs}
\usepackage[ruled,vlined]{algorithm2e}
\let\cline\cmidrule
\newcommand\T{\rule{0pt}{1.5ex}}

\usepackage{color}
\makeatletter
\newcommand*\bigcdot{\mathpalette\bigcdot@{0.5}}
\newcommand*\bigcdot@[2]{\mathbin{\vcenter{\hbox{\scalebox{#2}{$\m@th#1\bullet$}}}}}

\makeatother

\usepackage{amssymb}

\title{Reserve Allocation in Active Distribution Systems for Tertiary Frequency Regulation: A Coalitional Game Theory-based Approach}

\author{Mukesh Gautam, \emph{Student Member, IEEE}, Mohammad MansourLakouraj, \emph{Student Member, IEEE}, \\Rakib Hossain, \emph{Student Member, IEEE}, Narayan Bhusal, \emph{Member, IEEE},\\ Mohammed Benidris, \emph{Senior Member, IEEE}, and Hanif Livani, \emph{Senior Member, IEEE}\\ 
Department of Electrical \& Biomedical Engineering, University of Nevada, Reno \\
(emails: \{mukesh.gautam, mansour, rhossain, bhusalnarayan62\}@nevada.unr.edu, \{mbenidris, hlivani\}@unr.edu)\vspace{-0.2ex}}

\begin{document}
\maketitle

\begin{abstract}
Participation of distribution systems in frequency regulation has become an important factor for grid operation after the integration of distributed energy resources (DERs). Moreover, electricity regulatory authorities (e.g., Federal Electricity Regulatory Commission in the United States) are recommending that DERs should participate in energy and reserve markets, and a mechanism needs to be developed to facilitate DERs’ participation at the distribution level. Although the available reserve from a single distribution system may not be sufficient for tertiary frequency regulation, stacked reserves from many distribution systems can enable them to participate in tertiary frequency regulation at scale. This paper proposes a coalitional game theory-based approach for reserve optimization to enable DERs participate in tertiary frequency regulation. A two-stage approach is proposed to effectively and precisely allocate spinning reserve requirements from each DER in distribution systems. In the first stage, two types of characteristic functions: worthiness index (WI) and power loss reduction (PLR) of each coalition are computed. In the second stage, the equivalent Shapley values are computed based on the characteristic functions, which are used to determine distribution factors for reserve allocation among DERs. The effectiveness of the proposed method for allocation of reserves and determination of active power set-points of DERs is demonstrated through case studies on IEEE 13-node, IEEE 34-node, and IEEE 123-node distribution test systems.
\end{abstract}
\begin{IEEEkeywords}
 Coalitional game theory, distributed energy resources, tertiary frequency regulation, Shapley Value.
\end{IEEEkeywords}

\section*{Nomenclature}
\begin{tabbing}
AAAAAA \= AAAAAAAAAAAAAAAAAAAAAAAAAAAAAAAA \kill 
$\mathscr{N}$ \> players' set of a coalitional game\\
$S$ \> a coalition that is subset of $\mathscr{N}$\\
$2^{\mathscr{N}}$ \> possible coalitions' set \\
$S\backslash\{j\}$ \> coalition set that excludes player $j$\\
$\psi_j$ \> Shapley value of player $j$ \\
$n$ \> number of DERs or players \\
$WI_i$ \> worthiness index of the $i$\textsuperscript{th} DER \\
$PI_i$ \> performance index of the $i$\textsuperscript{th} DER \\

$PCAR_i$ \> priced capacity available for reserve \\ \>of the $i$\textsuperscript{th} DER \\
$UCAR_i$ \> unpriced capacity available for reserve \\ \>of the $i$\textsuperscript{th} DER \\
$TUCAR$ \> total unpriced capacity available for reserve \\
$P_{ci}$ \> sellable capacity of $i$\textsuperscript{th} DER \\
$P_{ei}$ \> market clearing capacity of $i$\textsuperscript{th} DER \\
$P_R$ \> total power reserve to be allocated from \\ \>a distribution system \\
$\psi_{1,i}$ \> Shapley value of $i$\textsuperscript{th} DER corresponding to \\ \>the first characteristic function \\
$\psi_{2,i}$ \> Shapley value of $i$\textsuperscript{th} DER corresponding to \\ \>the second characteristic function \\
$\psi^{norm}_{1,i}$ \> normalized Shapley value of $i$\textsuperscript{th} DER \\ \>corresponding to the first characteristic function \\
$\psi^{norm}_{2,i}$ \> normalized Shapley value of $i$\textsuperscript{th} DER \\ \>corresponding to the second characteristic function \\
$\psi^{eqv}_i$ \> equivalent Shapley value of $i$\textsuperscript{th} DER \\
$DF_i$ \> distribution factor of $i$\textsuperscript{th} DER \\
$P_{ri}$ \> active power set-point of $i$\textsuperscript{th} DER after \\ \>reserve allocation\\
$RBP_i$ \> reserve bid price of $i$\textsuperscript{th} DER \\
$R^2_i$ \> coefficient of determination \\ \>or R-squared score of $i$\textsuperscript{th} DER \\
$\alpha_c$ \> critical load factor \\
\end{tabbing}
\section{Introduction}
Integration of distributed energy resources (DERs) has brought several challenges as well as benefits to power grid operation including frequency control and regulation \cite{rapizza2020fast}. Frequency regulation problems (e.g., frequency deviation) occur when there is an imbalance between the generation and load, which can happen due to several factors including faults, large load changes, generating unit tripping, and islanding parts of the grid \cite{zhou2019optimal}. Frequency regulating devices are used in these situations to correct for frequency deviations. Under vertically integrated monopolistic structure of utilities, system operators set operating points of individual generators based on an optimal power flow (OPF) solution, which minimizes the overall operating cost of generation subjected to network and reserve constraints. On the other hand, in deregulated power systems, the main function of tertiary frequency regulation schemes is to maximize the net social welfare through allocating adequate spinning reserve from generators or DERs participating in primary and secondary frequency regulation \cite{machowski2020power}. Although the contribution of a single DER in frequency control and regulation is not significant, the accumulated contribution from a fleet of DERs can enable them to collectively participate in frequency control and regulation. However, allocating impactful reserves from DERs is a challenging task and requires flexible and efficient solutions.

Frequency regulation and control in power systems are generally categorized into primary, secondary, and tertiary regulation/control. A primary frequency control is the first line of defense that responds to a disturbance quickly (within milliseconds). If primary frequency control is unable to fully compensate for frequency deviations due to the disturbance, secondary frequency control schemes kick in, adjusting the power generated by the generating units to reinstate frequency to its nominal value. After executing the secondary frequency control, the resulting operating points of different generators may not be economically optimal. The tertiary frequency control schemes can then be initiated by system operators to minimize cost of generation through solving an optimal power flow while satisfying the reserve requirement. The tertiary frequency control/regulation schemes used in traditional power systems, on the other hand, may not be directly applicable to DER-based distribution networks. In order for distribution systems to participate in tertiary frequency regulation, an efficient approach for allocating reserves among DERs and determining their active power set points is required.  

Several approaches for tertiary frequency regulation and control in transmission systems and microgrids have been reported in the literature. An approach for optimal tertiary frequency control has been proposed in \cite{perninge2017optimal}, which also considers regulation based on electricity market.
A model predictive control (MPC)-based approach has been proposed in \cite{abbaspourtorbati2012towards} for the activation of tertiary frequency control reserves. A mixed integer linear programming-based optimization tool has been proposed in \cite{malik2012decision} for the activation of tertiary frequency control reserves at the transmission level. In \cite{delfino2002load}, a load frequency control has been proposed from the perspective of restructured power systems. In \cite{donde2001simulation}, the traditional automatic generation control (AGC) has been modified and implemented from the perspective of deregulated power systems. In \cite{bovera2021data}, a data-driven approach for the estimation of secondary and tertiary reserves has been presented and tested on a real system.
Although several methods and algorithms have been developed and employed for tertiary frequency regulation for microgrids and transmission systems, enabling active distribution systems to provide reserve is still a challenge.

Applications of coalitional game theoretic approaches in power and energy systems have gained significant momentum in recent years due to their ability to uniquely assign payoffs among players of the game taking into consideration their marginal contributions. A coalitional game theory-based energy management system has been proposed in \cite{querini2020cooperative} to facilitate power exchange of microgrids connected with the main grid. In \cite{gautam2022coop_DER}, a coalitional game theory-based approach has been proposed to find optimal sizes and locations of distributed energy resources. A coalitional game theory-based approach for participation of active distribution system in secondary frequency regulation has been proposed in \cite{gautam2022coop_SRF}.
In \cite{nazari2021economy}, a game-theoretic approach based on computation of the locational marginal price at each bus has been proposed for reliability enhancement and power loss reduction in active distribution systems and microgrids, where each player of the game receives economic incentives when system reliability is improved and power loss is reduced.    

This paper proposes a coalitional game theoretic two-stage approach to allocate reserves among DERs for tertiary frequency regulation. In the first stage, two types of characteristic functions: worthiness index (WI) and power loss reduction (PLR) are computed for each set of possible coalitions of participating DERs. The WI is an indicator of the worth or value of DERs for maximization of social welfare and the PLR denotes reduction in power loss as a result of DER addition. Equivalent Shapley values and hence distribution factors of DERs are determined in the second stage, which are utilized for the allocation of reserves among DERs and determination of their active power set-points. The effectiveness and efficacy of the proposed approach is demonstrated through case studies on several test systems.
The main contributions of this paper are as follows:
\begin{itemize}
    \item Proposing a framework for participation of DERs in tertiary frequency regulation. The proposed framework will help system operators to allocate reserve among different DERs in an effective and efficient manner.
    \item Developing a coalitional game theoretic framework to allocate reserve among different DERs for tertiary frequency regulation. The coalitional game theoretic approaches have the ability to uniquely assign payoffs among players of the game.
    \item Proposing two types of characteristic functions (WI and PLR) of each coalition for the coalitional game model under consideration. In order to maximize the social welfare or benefit, three factors priced capacity available for reserves (PCAR), reserve bid price (RBP), and performance index (PI) are proposed to compute the WI of each DER.
    \item Proposing a distribution factor based on the Shapley value to allocate the reserves among different DERs. The Shapley value considers the marginal contribution of each DER for reserve allocation among participating DERs. The active power set points of each DER is then calculated using the proposed distribution factor.
\end{itemize}

The remainder of the paper is laid out as follows. The formulation of tertiary frequency regulation as a coalitional game is presented in Section \ref{formulation}.
Coalitional game theory, including the Shapley value, is described in Section \ref{game}. The proposed approach for tertiary frequency regulation is explained in Section \ref{methodology}. Case studies on the modified IEEE 13-node, IEEE 34-node, and IEEE 123-node distribution test systems are described in Section \ref{cases}. Finally, in Section \ref{conclusion}, there are some concluding remarks.

\section{Tertiary Frequency Regulation and Coalitional Game} \label{formulation}
Frequency fluctuation in a power system occurs when there is a mismatch between electricity generation and consumption. Under such scenarios, frequency control/regulation devices come into play to bring the system frequency back to its nominal value and ensure that it stays within a specified range of the nominal frequency \cite{rebours2007survey, gautam2020sensitivity}. Conventionally, the frequency control is usually categorized into primary, secondary, and tertiary frequency control. The primary frequency control refers to the localized control of a generator's power according to its droop control characteristic. The aim of the primary frequency control is to re-establish the balance between generation and consumption (load plus losses) in the system at a frequency that may be different from the nominal frequency \cite{rebours2007survey}---i.e., the primary frequency control may stabilize the frequency but it may not be able to restore it to the nominal value. 
Whereas the primary frequency control simply tries to maintain system stability by ensuring balance between generation and load, the secondary frequency control restores the frequency to the nominal value within seconds to a few minutes \cite{dorfler2015breaking, gautam2021cooperative_PESGM}.
\begin{figure}
    \centering
    \includegraphics[scale=0.8]{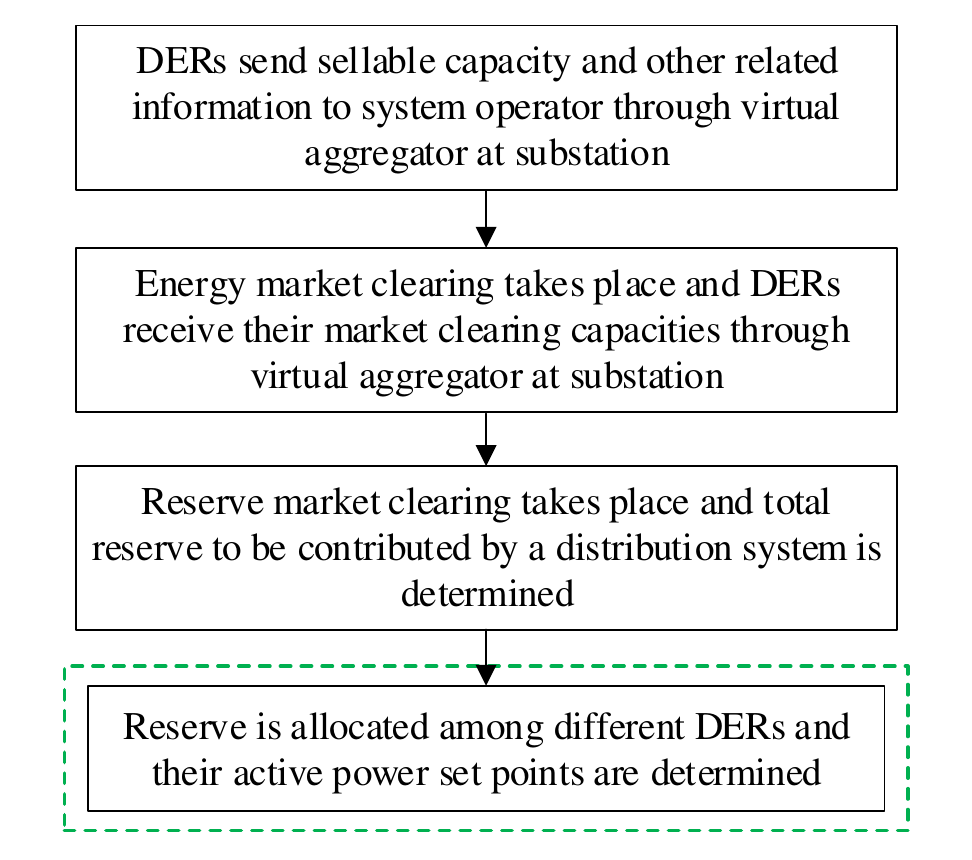}
    \caption{Layout of the overall process of reserve optimization}
    \label{fig:overall}
\end{figure}
\begin{figure}
    \centering
    \includegraphics[scale=0.9]{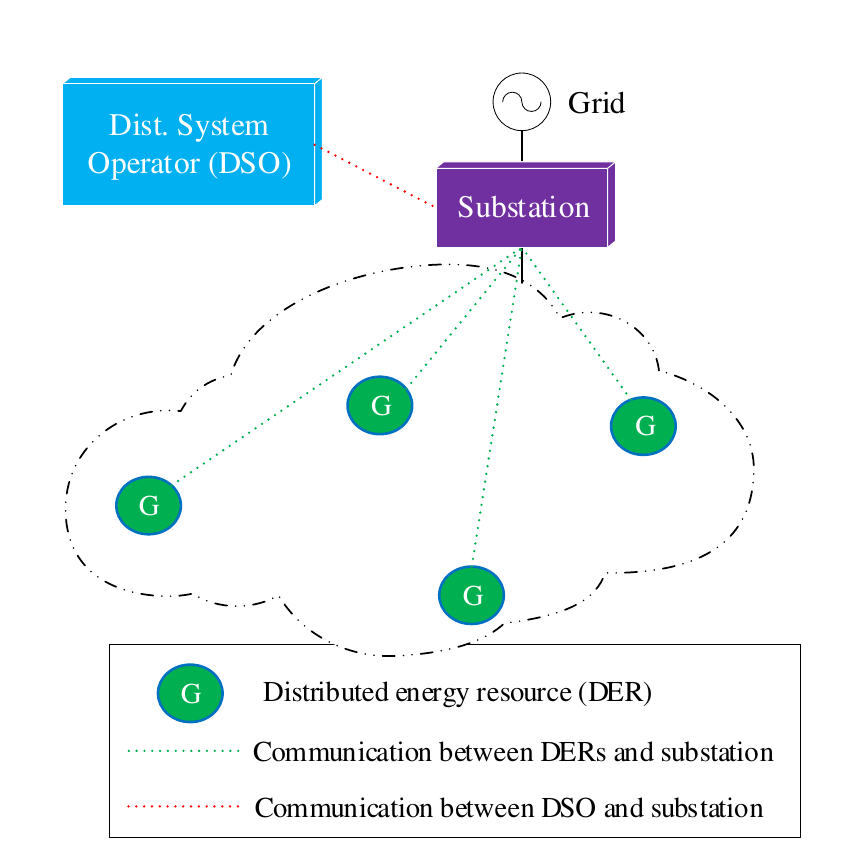}
    \vspace{0ex}
    \caption{Layout of the proposed coalitional game theoretic approach}
    \label{fig:coop_game}
\end{figure}

The tertiary frequency regulation scheme economically allocates reserves considering the technical constraints for frequency control and response. If the reserve or part of the reserve is used by the secondary frequency control after a disturbance or imbalance between generation and load, the tertiary frequency regulation restores the reserve to the desired level. In other words, the tertiary frequency control is used to relieve secondary reserve, following the loss of large generator or load, or to correct prediction inaccuracies \cite{rebours2007survey}. The tertiary frequency control is triggered manually in response to system operators' requests.

Under vertically integrated monopolistic structure of power systems, the tertiary frequency regulation is responsible for optimal power sharing among the generators to minimize the overall cost or to maximize the net social welfare \cite{machowski2020power}. In case of electric utilities operating under partial or fully deregulated environment, the operating reserve capacity is treated as a commodity like energy in an electricity market \cite{wang2005operating}.
The co-optimization of energy and reserve dispatch has been investigated in \cite{gan2003energy} in the setting of a pool-based market, where the modeling of lost opportunity has also been presented using a general approach. An offering mechanism of aggregators in the energy and reserve power markets has been investigated in \cite{baringo2021offering}. 
A new formulation of independent clearing of energy and reserve market has been introduced in \cite{afshar2008method}, which also considers lost opportunity cost of generating units for clearing reserve market. The co-optimization of energy and reserve has been proposed in \cite{pavic2019energy} with the consideration of two compensation mechanisms: uplift payments and lost opportunity costs. A mathematical model for battery energy storage system (BESS) participation in reserve market has been developed in \cite{padmanabhan2019battery}. A bidding strategy for the participation of BESS in the reserve market has been presented in \cite{merten2020bidding}.

In this paper, the task of allocating reserves and determining DERs' power set-points is regarded as a coalitional game and the participating DERs are regarded as players of the game. As a motivation for DERs to participate in tertiary frequency regulation, DERs are allowed to send their bid prices for reserves. Apart from this, DERs also send information related to their sellable capacity to a virtual aggregator at the substation, which then sends this information to system operators.
Figure~\ref{fig:overall} shows the layout of the overall process of reserve optimization. The focus of this paper is the last step of the process (shown inside green dashed rectangle). 
Figure~\ref{fig:coop_game} shows the layout of the proposed coalitional game theoretic approach to allocate reserves in an active distribution system for the tertiary frequency regulation. Based on the information received from DERs and their performance history, the distribution system operator computes two types of characteristic functions: WI and PLR of each coalition for fair allocation of reserves among DERs. 

\section{Coalitional Game Theory and Shapley Value}\label{game}
In game theory, games are generally categorized into two groups: (a) coalitional games and (b) non-coalitional games. In non-coalitional games, there is no coalition or cooperation between players and they compete among each other to maximize their individual utilities, while in coalitional games the players can establish alliances or coalitions with one another to maximize coalitional and individual utilities. Since players form coalitions to maximize their individual utilities, a coalition must always result in utilities that are equal to or larger than the individual player's utilities. Non-coalitional games focus mainly on maximizing individual utilities of the players, while coalitional games focus on improving joint utility of the coalition \cite{gautam2022cooperative}. Since the frequency of the system needs to be regulated through combined effort of all DERs, coalitional games are more suitable candidates for the tertiary frequency regulation problem under consideration; therefore, hereafter only coalitional games are discussed. A coalitional game is defined by assigning a value to each of the coalitions. The coalitional game is composed of the following two components:
\begin{enumerate}
    \item A finite players' set $\mathscr{N}$, known as the grand coalition.
    \item A characteristic function $V(S):2^{\mathscr{N}} \rightarrow \mathbb{R}$ that maps the set of all possible player coalitions to a set of coalitional values that satisfy $V(\phi)=0$.
\end{enumerate}

The characteristic function, representing worth or value of each coalition, is defined in every coalitional game.
The characteristic function of a coalition is the aggregated worth of all coalition members. Solution paradigms such as the Shapley value, the core, the Nucleolus, and the Nash-bargaining solution are used to allocate the overall payout or incentive among individual players of a coalitional game. 

\subsection{The Core of a Coalitional Game}
 In game theory, the core is the set of possible assignments that cannot be enhanced more through any alternative coalitions. The core is a set of payout assignments that ensures no player or player group has a motivation to quit $\mathscr{N}$ to establish a new coalition. Mathematically, the core is defined as follows \cite{shapley1975competitive}. 
 \begin{equation}
     \mathcal{C} = \left\{ \alpha: \sum_{j\in \mathscr{N}} \alpha_j =V(\mathscr{N}) \text{ and } \sum_{j\in S} \alpha_j \geq V(S), \forall S \subset \mathscr{N} \right\}
 \end{equation}

\subsection{The Shapley Value}
The Shapley value is a solution paradigm of coalitional game theory. The Shapley value is an approach to allocate the overall earnings to individual players when all participants participate in the game.
The Shapley value of a coalitional game is expressed as follows \cite{curiel2013cooperative}.
\begin{equation}
    \psi_j(V) =\hspace{-1.5ex} \sum_{S \in 2^{\mathscr{N}}, j \in S}\hspace{-1.5ex} \frac{(\lvert S\rvert-1)!(n-\lvert S\rvert)!}{n!}[V(S)-V(S\backslash\{j\})]
    \label{eqn:shapley}
\end{equation}
where $n=\lvert \mathscr{N} \rvert$ denotes the total number of players. 

The Shapley value has a number of important properties, which are listed below:
\begin{enumerate}
    \item \textit{Efficiency:} The grand coalition's value is equal to the total of all players' Shapley values, therefore all gains are allocated among the players. Mathematically,
    \begin{equation}
        \sum_{j \in \mathscr{N}} \psi_j(V) = V(\mathscr{N})
    \end{equation}
    \item \textit{Individual Rationality:} When a player participates in a coalition, then its Shapley value should exceed its individual value. Mathematically,
    \begin{equation}
       \psi_j(V) \geq V(\{j\}), \forall j \in \mathscr{N}
    \end{equation}
    \item \textit{Symmetricity:} When two players in a coalition make the same contribution, their Shapley values must be equal. Mathematically, for two players $i$ and $j$ satisfying $V(S \cup \{i\}) = V(S \cup \{j\})$ for each coalition $S$ without $i$ and $j$,
    \begin{equation}
       \psi_i(V) = \psi_j(V) 
    \end{equation}
    \item \textit{Dumminess:} When a player does not increase the coalition's value, its Shapley value should be zero. Mathematically, for player $i$ satisfying $V(S) = V(S \cup \{i\})$ for each coalition $S$ without $i$,
    \begin{equation}
       \psi_i(V) = 0 
    \end{equation}
    \item \textit{Linearity:} The Shapley value corresponding to the sum of characteristic functions equals the sum of Shapley values corresponding to the individual characteristic functions. Mathematically, for two characteristic functions $V_1$ and $V_2$ of a coalitional game,
    \begin{equation}
        \psi(V_1+V_2) = \psi(V_1) + \psi(V_2)
    \end{equation}
\end{enumerate}

\section{The Proposed Approach}\label{methodology}
In this section, an approach for reserve allocation among DERs in a distribution system is developed using a coalitional game theoretic approach. There are two stages to the proposed coalitional game theoretic approach. In the first stage, the two types of characteristic functions: WI and PLR are computed. Shapley values are calculated and then reserves are allocated in the second stage. Moreover, we have proposed various metrics including priced capacity available for reserve (PCAR), unpriced capacity available for reserve (UCAR), and Distribution Factor (DF), which are explained in this section. These metrics are pivotal for fair allocation of reserves among DERs.

The WI of each DER indicates the worth or value of each DER for the allocation of reserve so as to maximize the social welfare or benefit. The three factors: PCAR, reserve bid price (RBP), and performance index (PI) are taken into consideration for computing WI of each DER.

The available capacity for reserve (i.e., after energy market clearance) from the $i$\textsuperscript{th} DER can be calculated as follows.
\begin{equation}
    \mbox{Available Capacity for Reserve} = P_{ci} - P_{ei} \label{eqn:availreserv}
\end{equation}
where $P_{ci}$ is total capacity of the $i$\textsuperscript{th} DER available for selling and $P_{ei}$ is the accepted capacity of the $i$\textsuperscript{th} DER after clearing energy market.

Since the $i$\textsuperscript{th} DER has accepted $P_{ei}$ through market clearing and its local demand is already met, this reserve can be considered unpriced and is, therefore, named as unpriced capacity available for reserve (UCAR). However, it is to be noted that the DER will get paid whenever the reserve is used for secondary frequency control. The total unpriced capacity available for reserve (TUCAR) of a distribution system is calculated by adding UCAR of each DER behind the substation of the distribution system.

\begin{equation}
    TUCAR = \sum_{i=1}^{n} UCAR_i = \sum_{i=1}^n P_{ci}-P_{ei} \label{eqn:tucar}
\end{equation}
\begin{figure}
    \centering
    \includegraphics[scale=0.85]{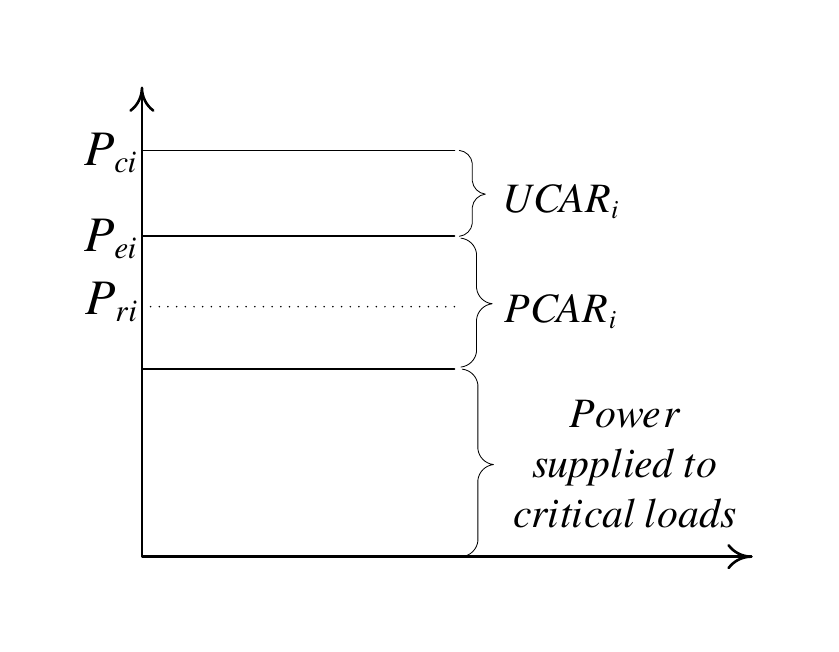}
    \vspace{-3ex}
    \caption{Allocation of reserves for $i$\textsuperscript{th} DER}
    \label{fig:Reserves_DERi}
\end{figure}
If the reserve command received from the distribution system operator, $P_R$, is higher than TUCAR, then an additional reserve needs to be allocated from $P_{ei}$. Under such circumstances, the DER has to be paid lost opportunity cost for the additional reserve. Thus, the capacity available for an additional reserve after deducting the power being supplied to critical loads is named as priced capacity available for reserve (PCAR) and is computed as follows.
\begin{equation}
    PCAR_i = (1-\alpha_c) \times P_{ei} \mbox{,} \label{eqn:pcar}
\end{equation}
where $\alpha_c$ is the critical load factor denoting the fraction of power being supplied by each DER to critical loads. This fraction of power cannot be allocated for reserves.

For the case of $P_R$ being higher than TUCAR, one of the main functions of tertiary frequency controller or regulator is to allocate $P_R-TUCAR$ optimally among the participating DERs. 

The worthiness index (WI) of the $i$\textsuperscript{th} DER is defined as follows.
\begin{equation}
    WI_i = \frac{PCAR_i}{RBP_i}\times PI_i \mbox{,} \label{eqn:wi}
\end{equation}
where $PCAR_i$ is PCAR of the $i$\textsuperscript{th} DER, $RBP_i$ is reserve bid price of the $i$\textsuperscript{th} DER, and $PI_i$ is the performance index of the $i$\textsuperscript{th} DER.

The performance history consisting of the values of committed power and supplied
power of DERs is utilized to determine the performance index (PI) of each DER. The errors between committed power and supplied power are utilized in order to compute the coefficient of determination (also referred to as R-squared score), which is, here, the performance index (PI) of each DER.
The expression for the performance index of the $i$\textsuperscript{th} DER is as follows.
\begin{equation}
    PI_i = R_i^2 = 1- \frac{ESS_i}{TSS_i} \mbox{,}\label{eqn:pi}
\end{equation}
where $R_i^2$ is the R-squared score of the $i$\textsuperscript{th} DER, $ESS_i$ is the sum of squares of errors between committed power and supplied power of the $i$\textsuperscript{th} DER, and $TSS_i$ is the total sum of squares of committed power of the $i$\textsuperscript{th} DER.

WI of each DER acts as the first characteristic function of the proposed coalitional game and PLR acts as the second characteristic function. PLR is the difference between the active power loss of the system with DERs of a particular coalition and that without any DER.

Based on each type of characteristic function of participating DERs and their coalitions, Shapley values ($\psi_{1,i}$ and $\psi_{2,i}$) are calculated using \eqref{eqn:shapley}. The normalized Shapley values corresponding to each characteristic function are then calculated as follows.
\begin{equation} 
     \psi_{1,i}^{norm} =\psi_{1,i} \Big/ \sum_{k=1}^{n}\psi_{1,k} \mbox{,} \label{eqn:norm1}
\end{equation}
\begin{equation} 
     \psi_{2,i}^{norm} =\psi_{2,i} \Big/ \sum_{k=1}^{n}\psi_{2,k} \mbox{.} \label{eqn:norm2}
\end{equation}

Using normalized Shapley values calculated in \eqref{eqn:norm1} and \eqref{eqn:norm2}, the equivalent Shapley value of the $i$\textsuperscript{th} DER is computed as follows.
\begin{equation}
    \psi_i^{eqv} = \frac{\psi_{1,i}^{norm}+\psi_{2,i}^{norm}}{2} \label{eqn:eqShap}
\end{equation}

The distribution factor (DF) of the $i$\textsuperscript{th} DER is then calculated using \eqref{eqn:DF}. The distribution factors are utilized to distribute the difference of $P_R$ and $TUCAR$ (i.e., $P_R$ -- $TUCAR$) among the participating DERs.
\begin{equation} 
     DF_i =\psi_i^{eqv} \Big/ \sum_{k=1}^{n}\psi_k^{eqv} \label{eqn:DF}
\end{equation}

The active power set point of the $i$\textsuperscript{th} DER is then determined as follows.
\begin{equation}
    P_{ri} = P_{ei}-(P_R - TUCAR) \times DF_i \label{eqn:Pri}
\end{equation}

The proposed approach to compute reserves allocated and set points of DERs for tertiary frequency regulation can be summarized as follows.
\begin{enumerate}
    \item Gather information on lines, loads, DERs, etc. in the system.
    \item Read $P_c$, $P_e$, and RBP of each DER.
    \item Determine the performance index (PI) of each DER based on past history of committed power and supplied power.
    \item Determine priced capacity available for reserve (PCAR) and worthiness index (WI) of each DER.
    \item Compute two types of characteristic functions: worthiness index (WI) and power loss reduction (PLR) of each coalition by enumerating all potential DER coalitions.
    \item Compute Shapley values using \eqref{eqn:shapley} and normalized Shapley values using \eqref{eqn:norm1} and \eqref{eqn:norm2}.
    \item Compute the equivalent Shapley values using \eqref{eqn:eqShap} and the distribution factors using \eqref{eqn:DF}.
    \item Determine the allocated reserves and set-points of DERs using \eqref{eqn:Pri}.
\end{enumerate}

\begin{figure}[h!]
    \centering
    \includegraphics[scale=0.75]{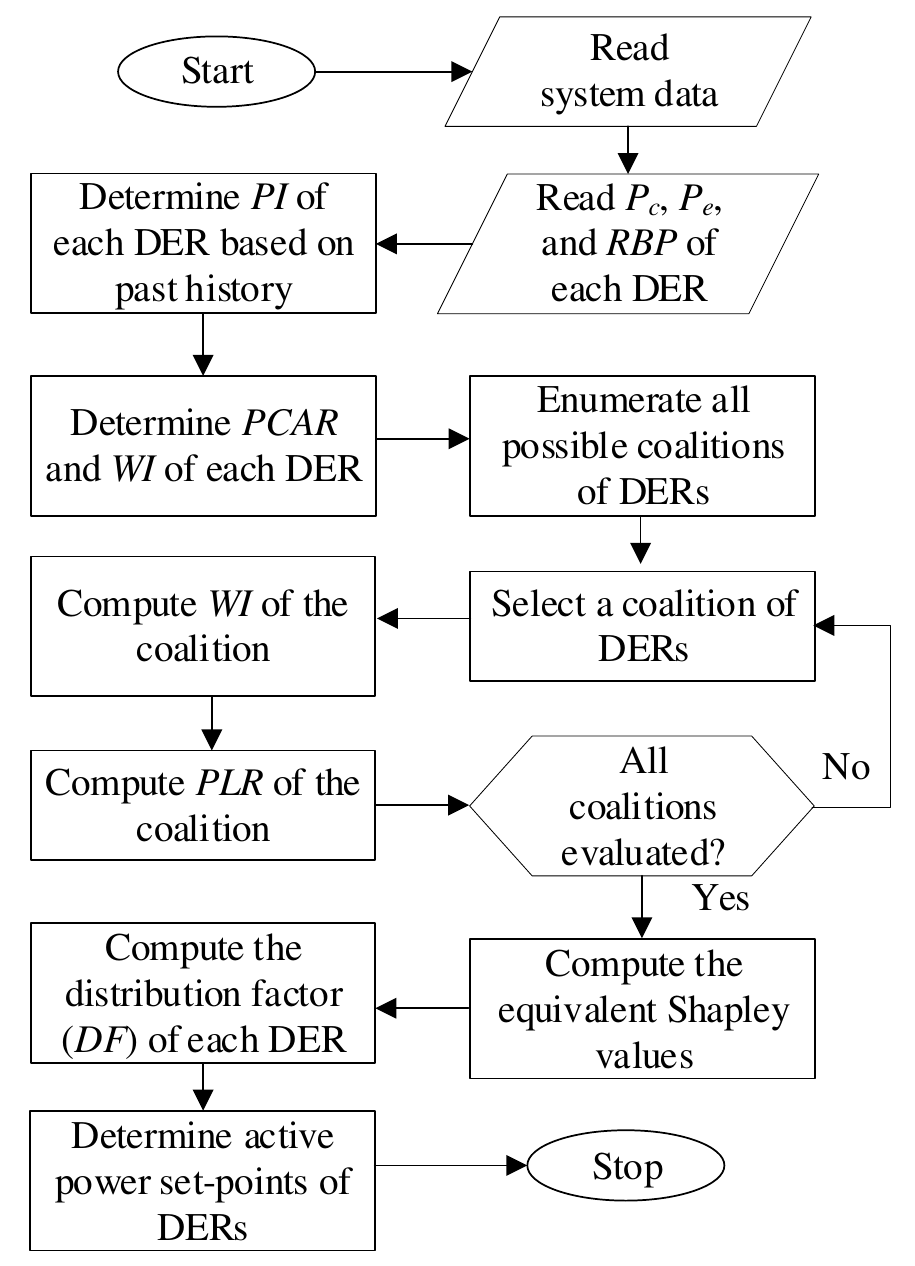}
    \caption{Flowchart of the proposed approach}
    \label{fig:flow_chart}
\end{figure}

The flowchart of the proposed approach is shown in Figure~\ref{fig:flow_chart}.

\section{Case Studies and Discussions}\label{cases}
To demonstrate the effectiveness and efficacy of the proposed approach and the solution algorithm, modified IEEE 13-node, IEEE 34-node, and IEEE 123-node distribution test systems are used for numerical simulations.
\subsection{System Descriptions}
\begin{figure}[h!]
    \centering
    \includegraphics[scale=0.51]{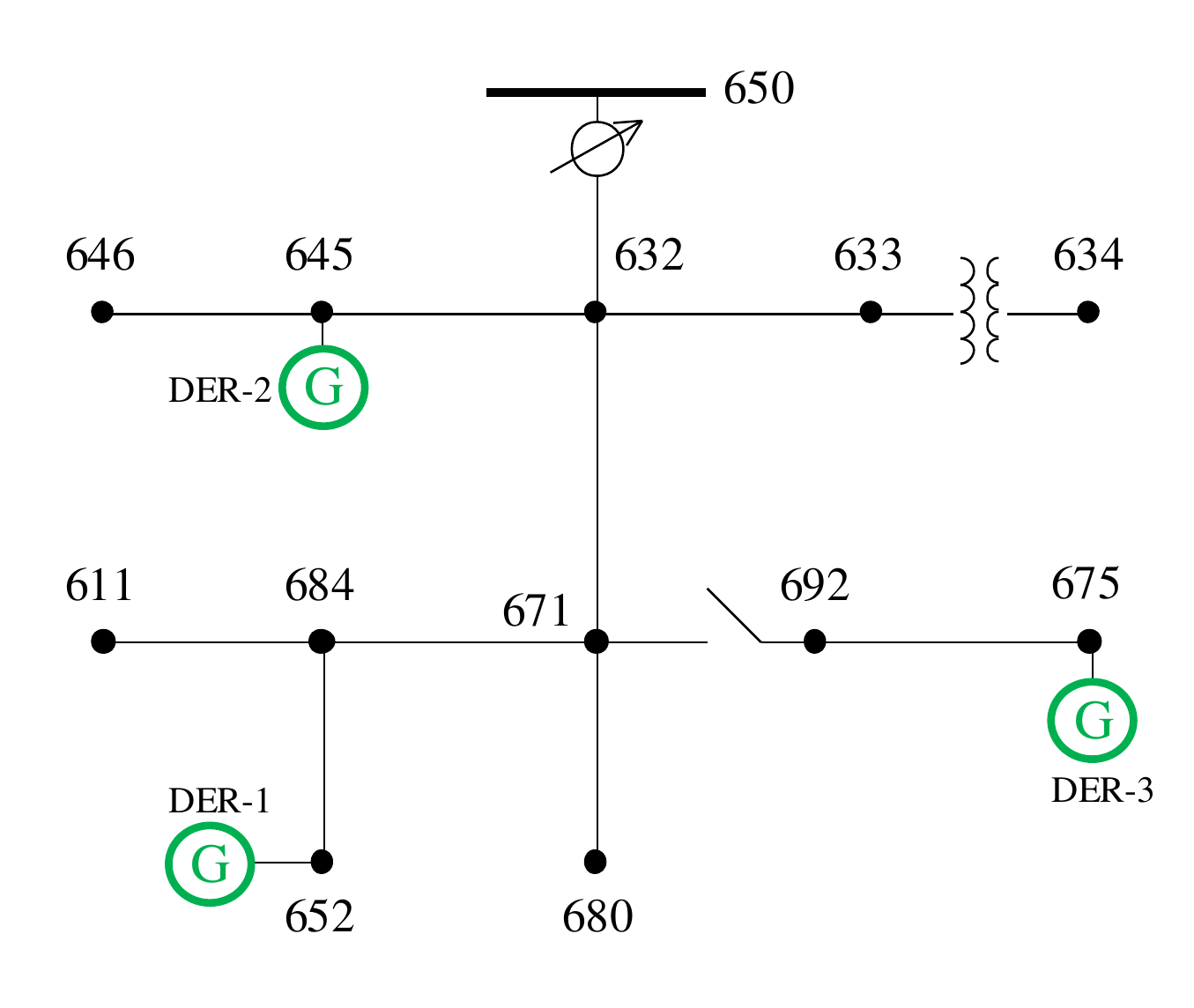}
    \caption{The modified IEEE 13-node distribution system}
    \label{fig:13node}
\end{figure}

\begin{table}[]
\vspace{-2ex}
\caption{Different parameters of DERs in the case of the modified IEEE 13-node system}
\centering
\begin{tabular}{|c|c|c|c|c|c|c|}
\hline
DERs & Locations & $P_c$  & $P_e$   & $PCAR$ & $RBP$  &  $PI$     \\ 
 & (nodes) & (kW) & (kW)& (kW)  & (\$/kW) &  \\ \hline \hline
1 & 652 & 250 & 220 & 110 & 12 & 0.9998 \\ \hline
2 & 645 & 350 & 345 & 172.5 & 15 & 0.9009 \\ \hline
3 & 675 & 450 & 400 & 200 & 10 & 0.8272 \\ \hline
\end{tabular}
\label{tab:parameters13node}
\end{table}

The IEEE 13-node distribution test system is a 4.16 kV distribution system. It is a compact feeder with a high load capacity. A voltage regulator is connected to the substation, which is made up of three 1-phase units connected in a star configuration. Shunt capacitor banks, an in-line transformer, and unbalanced spot and distributed loads characterize the system. This system's total active and reactive loads are 3577 kW and 1725 kVAr, respectively \cite{IEEEFEEDERS}. Three DERs are added to this system, as indicated in Figure~\ref{fig:13node}.
\begin{figure}[h!]
    \centering
    \hspace{-4ex}
    \includegraphics[scale=0.65]{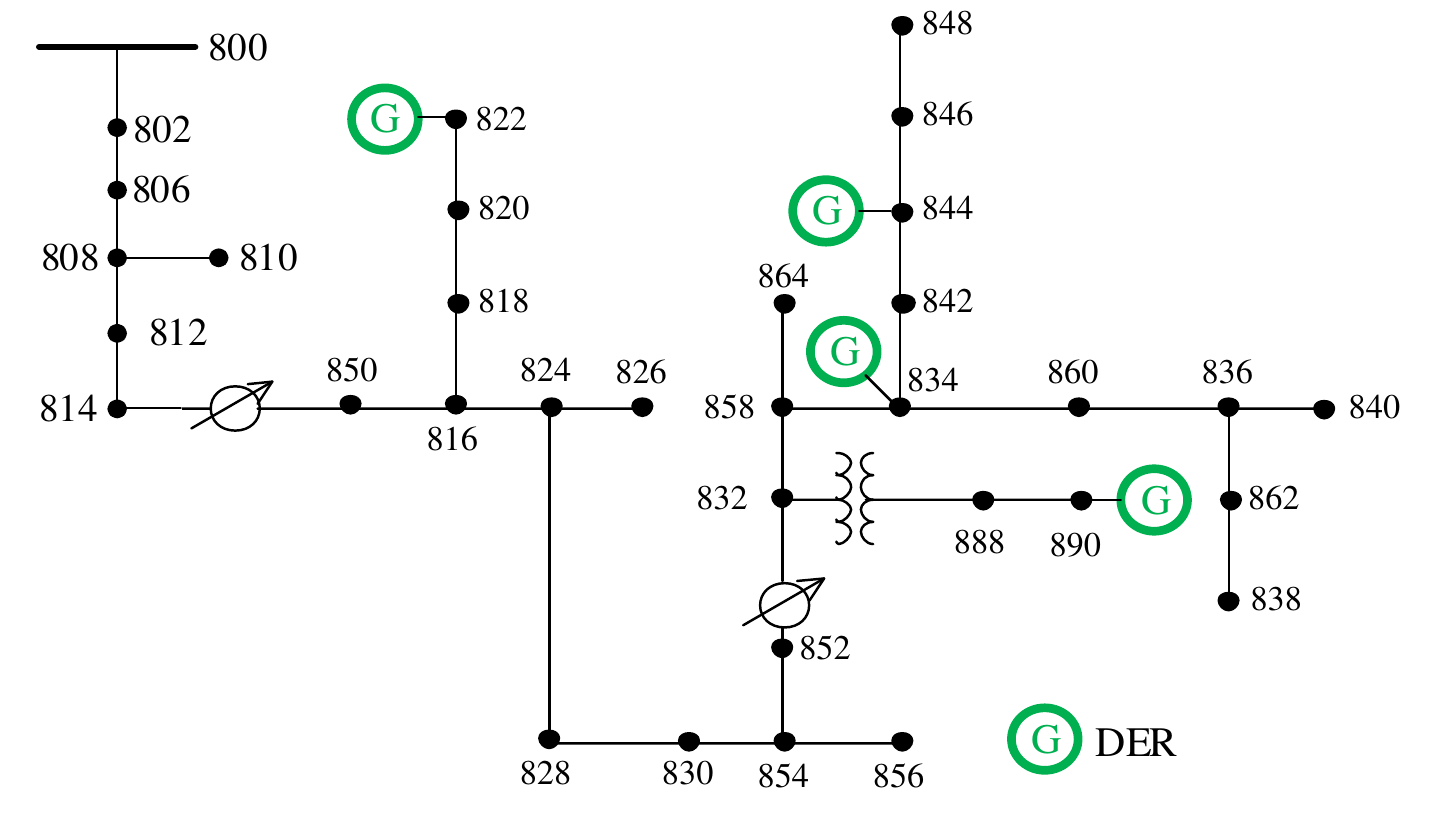}
    \vspace{-1.5ex}
    \caption{The modified IEEE 34-node distribution system}
    \label{fig:34node}
\end{figure}

The IEEE 34-node system is an actual distribution system in Arizona, USA. It's nominal voltage of 24.9 kV. It is a long, light-loaded feeder. Shunt capacitor banks, two in-line voltage regulators, an in-line transformer, and unbalanced spot and distributed loads characterize the system \cite{IEEEFEEDERS}. Three 3-phase DERs are added to this system, as indicated in Figure~\ref{fig:34node}, at nodes 844, 890, and 834, and a single-phase DER is added at phase 1 of node 822. 
\begin{table}[h!]
\caption{Different parameters of DERs in the case of the modified IEEE 34-node system}
\centering
\begin{tabular}{|c|c|c|c|c|c|c|}
\hline
DERs & Locations & $P_c$  & $P_e$   & $PCAR$ & $RBP$  &  $PI$     \\ 
 & (nodes) & (kW) & (kW)& (kW)  & (\$/kW) &  \\ \hline \hline
1 & 844 & 300 & 280 & 140 & 10 & 0.9999 \\ \hline
2 & 890 & 200 & 190 & 95 & 15 & 0.9732 \\ \hline
3 & 834 & 400 & 390 & 195 & 12 & 0.8863 \\ \hline
4 & 822 & 200 & 170 & 85 & 10 & 0.8335 \\ \hline
\end{tabular}
\label{tab:parameters34node}
\end{table}

\begin{figure*}
    \centering    \includegraphics[scale=1]{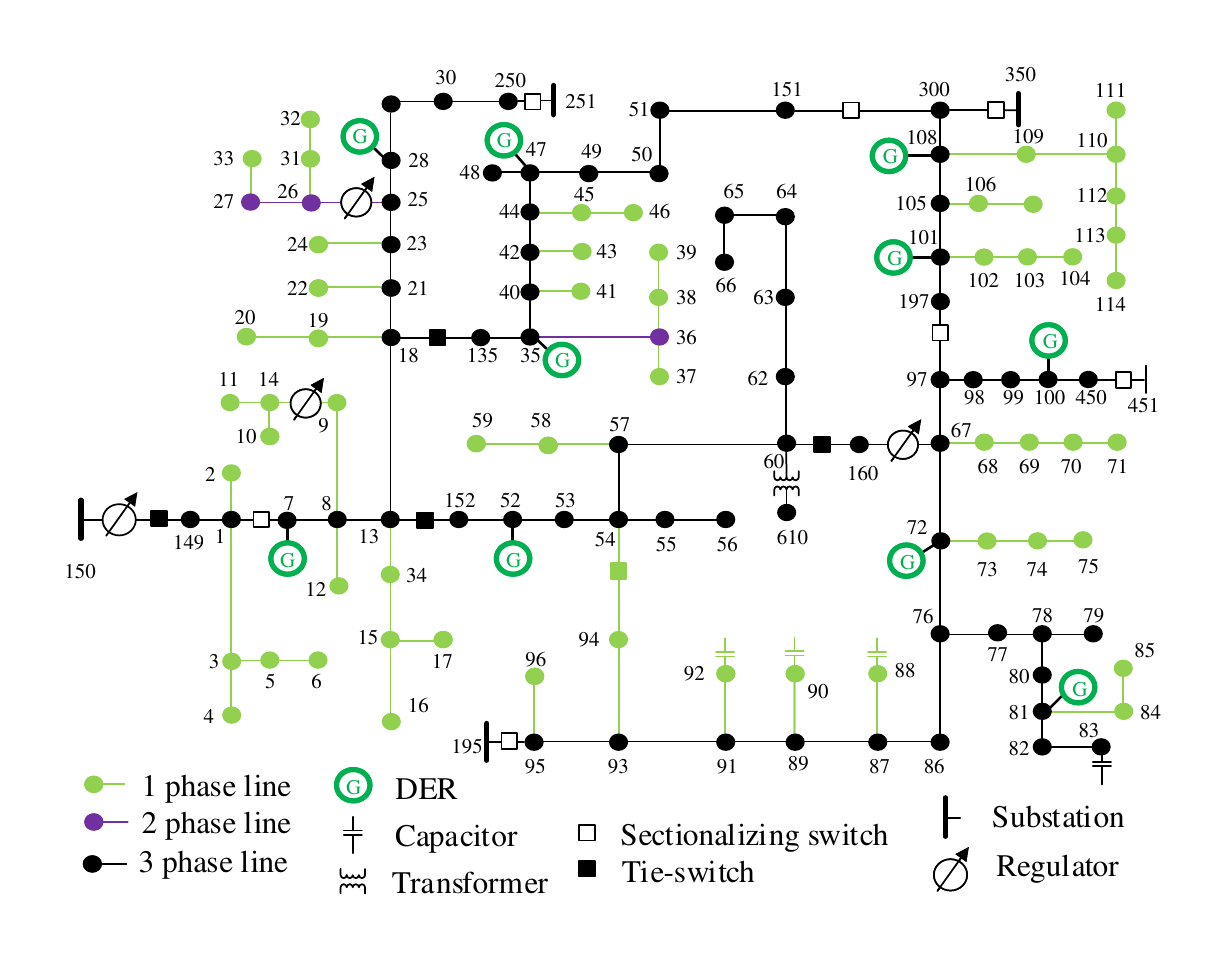}
    \caption{The modified IEEE 123-node distribution system}
    \label{fig:123node}
\end{figure*}

The IEEE-123 node system a 4.16 kV distribution system. The system includes overhead and underground line segments with varying phasing, shunt capacitor banks, four voltage regulators, and unbalanced spot loads with varied load types (PQ, constant I, constant Z). This system's total active and reactive loads are 3490 kW and 1925 kVAr, respectively \cite{IEEEFEEDERS}. Ten 3-phase DERs are added to this system at various nodes as shown in Figure~\ref{fig:123node}.

\begin{table}[h!]
\caption{Different parameters of DERs in the case of the modified IEEE 123-node system}
\centering
\begin{tabular}{|c|c|c|c|c|c|c|}
\hline
DERs & Locations & $P_c$  & $P_e$   & $PCAR$ & $RBP$  &  $PI$     \\ 
 & (nodes) & (kW) & (kW)& (kW)  & (\$/kW) &  \\ \hline \hline
1 & 7 & 100 & 90 & 45 & 10 & 0.9999 \\ \hline
2 & 28 & 100 & 95 & 47.5 & 15 & 0.9831 \\ \hline
3 & 35 & 110 & 100 & 50 & 12 & 0.9179 \\ \hline
4 & 47 & 80 & 70 & 35 & 10 & 0.7797 \\ \hline
5 & 52 & 100 & 95 & 47.5 & 15 & 0.9446 \\ \hline
6 & 72 & 100 & 90 & 45 & 11 & 0.8846 \\ \hline
7 & 81 & 150 & 140 & 70 & 12 & 0.9965 \\ \hline
8 & 100 & 100 & 100 & 50 & 15 & 0.9909 \\ \hline
9 & 101 & 70 & 60 & 30 & 10 & 0.9685 \\ \hline
10 & 108 & 80 & 70 & 35 & 12 & 0.9265 \\ \hline
\end{tabular}
\label{tab:parameters123node}
\end{table}

\begin{table}[]
\caption{Characteristic Functions: Worthiness Index (WI) and Power Loss Reduction for the Modified IEEE 13-node System\vspace{0.5ex}}
\centering
\begin{tabular}{|c|c|c|}
\hline
\hspace{3ex}Coalitions\hspace{3ex}& \hspace{3ex}Worthiness\hspace{3ex} & \hspace{3ex}Power Loss\hspace{3ex}   \\ 
of DERs & Index (WI) & Reduction (kW) \\\hline \hline
$1$ & $10.99$ & $8.59$    \\ \hline
$2$ & $7.88$ & $3.79$    \\ \hline
$3$ & $13.79$ & $15.54$    \\ \hline
$1,2$ & $18.88$ & $12.53$    \\ \hline
$1,3$ & $24.78$ & $21.87$    \\ \hline
$2,3$ & $21.67$ & $19.38$    \\ \hline
$1,2,3$ & $32.67$ & $25.91$    \\ 
 \hline
\end{tabular}
\label{tab:coalitions13}
\end{table}

\subsection{Parameters, Implementation, and Results}
The proposed approach starts by collecting information about sellable capacity ($P_c$), energy market clearing capacity ($P_e$), and reserve bid price (RBP) of each DER. Based on these values and critical load factor ($\alpha_c$), priced capacity available for reserve (PCAR) of each DER is calculated. The performance index (PI) of each DER is calculated based on past history of DERs consisting of committed power and supplied power. For the case of the modified IEEE 13-node system, the sellable capacities of DERs are 250 kW, 350 kW, and 450 kW. The energy market clearing capacities of DERs for a particular timestamp under consideration are 220 kW, 350 kW, and 400 kW. Reserve bid prices of DERs are \$10/kW, \$20/kW, and \$12/kW. These values are shown in Table~\ref{tab:parameters13node}. For a critical load factor ($\alpha_c=0.5$), PCAR of each DER obtained using \eqref{eqn:pcar} are 110 kW, 175 kW, and 200 kW, which are also shown in Table~\ref{tab:parameters13node}. The values of PI of each DER are 0.9998, 0.9009, and 0.8272, which are R-squared scores computed based on past history of committed power and supplied power of DERs. Based on \eqref{eqn:wi}, the PI of each DER has a direct impact on its worthiness index (WI). Since DER-3 has the least PI value, 0.8272, compared to DER-1, its WI (from Table~ \ref{tab:coalitions13}) is 13.79, which is higher than WI of DER-1 by only 2.8, even though its capacity is almost double.
Similarly, the different parameters of DERs in the case of the modified IEEE 34-node system and IEEE 123-node system are, respectively, shown in Table~\ref{tab:parameters34node} and Table~\ref{tab:parameters123node}.

\begin{table}[h!]
\caption{Characteristic Functions: Worthiness Index (WI) and Power Loss Reduction for the Modified IEEE 34-node System}
\centering
\begin{tabular}{|c|c|c|}
\hline
\begin{tabular}[c]{@{}c@{}}\hspace{2ex} Coalitions\hspace{2ex}  \\ of DERs\end{tabular} & \begin{tabular}[c]{@{}c@{}}\hspace{2ex}Worthiness\hspace{2ex} \\ Index (WI)\end{tabular} & \begin{tabular}[c]{@{}c@{}}\hspace{2ex}Power Loss\hspace{2ex} \\ Reduction (kW)\end{tabular} \\ \hline \hline
1   & 13.99 & 39.17   \\ \hline
2  & 6.16  & 42.89  \\ \hline
3   & 14.40  & 54.62  \\ \hline
4  & 7.08 & 17.00   \\ \hline
1, 2 & 20.16 & 79.53    \\ \hline
1, 3  & 28.40 & 88.68   \\ \hline
1, 4  & 21.08 & 56.69    \\ \hline
2, 3  & 20.56 & 92.92   \\ \hline
2, 4   & 13.25 & 58.66   \\ \hline
3, 4   & 21.49  & 70.08  \\ \hline
1, 2, 3  & 34.56  & 122.94   \\ \hline
1, 2, 4  & 27.24   & 94.74   \\ \hline
1, 3, 4  & 35.48  & 102.89 \\ \hline
2, 3, 4  & 27.65  & 106.88  \\ \hline
1, 2, 3, 4   & 41.65  & 136.17   \\ \hline
\end{tabular}
\label{tab:coalitions34}
\end{table}

As stated in Section~\ref{game}, a coalitional game is expressed in terms of a finite player set and characteristic functions. Two different types of characteristic functions (WI and PLR) are considered. These characteristic functions are defined for all possible sets of coalitions. The PLR for a coalition is the difference between power loss of the system without DERs and that with DERs of that coalition. For the case of the modified IEEE 13-node system, the possible sets of coalitions of DERs and the corresponding values of WI and PLR are shown in Table~\ref{tab:coalitions13}. As shown in the table, WI is 10.99 and PLR is 8.59 kW for DER-1. For DER-2, WI is 7.88 and PRL is 3.79 kW. For the coalition of DER 1 and 2, WI is 18.88 and PLR is 12.53 kW. Similarly, for the case of the modified IEEE 34-node system, the possible sets of coalitions and the corresponding values of WI and PLR are shown in Table~\ref{tab:coalitions34}. Similar table can be constructed for the case of the modified IEEE 123-node system which is not shown here.

\begin{table*}[]
\caption{Distribution Factors, Allocated Reserves, and Individual utilities of DERs for the 13-node System}
\centering
\begin{tabular}{|c|c|c|c|c|c|c|}
\hline
\multirow{2}{*}{DERs} & \multicolumn{3}{c|}{Proposed   Approach} & \multicolumn{3}{c|}{  Capacity-based Approach} \\ \cline{2-7}   & Dist. & Allocated  & Individual & Dist.       & Allocated  & Individual  \\ & Factors   &   Reserves (kW)  & utility &    Factors       &  Reserves (kW) & utility\\ \hline
1   & 0.2729 & 34.09 & 0.1965 & 0.2381 & 23.81 & 0  \\ \hline
2  & 0.2185  & 8.28 & 0.1265 & 0.3333  & 33.33 & 1.0939 \\ \hline
3 & 0.5086 & 57.63 & 0.1402 & 0.4286  & 42.86 & 0 \\ \hline
\end{tabular}
\vspace{-2.5ex}
\label{tab:Reserves_13}
\end{table*}

\begin{table*}[ht!]
\caption{Distribution Factors, Allocated Reserves, and Individual utilities of DERs for the 34-node System}
\centering
\begin{tabular}{|c|c|c|c|c|c|c|}
\hline
\multirow{2}{*}{DERs} & \multicolumn{3}{c|}{Proposed   Approach} & \multicolumn{3}{c|}{  Capacity-based Approach} \\ \cline{2-7}   & Dist. & Allocated  & Individual & Dist.       & Allocated  & Individual  \\ & Factors   &   Reserves (kW)  & utility &    Factors       &  Reserves (kW) & utility\\ \hline
1 & 0.2949 & 28.85 & 0.2949 & 0.2728  & 27.28 & 0.2424  \T \\ \hline
2  & 0.2141  & 16.42 & 0.4689 & 0.1818  & 18.18 & 0.5972   \T\\ \hline
3 & 0.3499  & 20.50 & 0.2791 & 0.3636   & 36.36 & 0.7010  \T\\ \hline
4 & 0.1410 & 34.23   & 0.1763 & 0.1818   & 18.18 & 0 \T\\ \hline
\end{tabular}
\label{tab:Reserves_34}
\end{table*}
\begin{table*}[ht!]
\caption{Distribution Factors, Allocated Reserves, and Individual utilities of DERs for the 123-node System}
\centering
\begin{tabular}{|c|c|c|c|c|c|c|}
\hline
\multirow{2}{*}{DERs} & \multicolumn{3}{c|}{Proposed   Approach} & \multicolumn{3}{c|}{  Capacity-based Approach} \\ \cline{2-7}   & Dist. & Allocated  & Individual & Dist.       & Allocated  & Individual  \\ & Factors   &   Reserves (kW)  & utility &    Factors       &  Reserves (kW) & utility\\ \hline
1 & 0.1798 & 13.59 & 0.3595 & 0.1010 & 10.10 & 0.0101  \T \\ \hline
2 & 0.2334  & 9.67 & 0.6884 & 0.1010  & 10.10  & 0.7522  \T\\ \hline
3 & 0.1810 & 13.62 & 0.3625 & 0.1111  & 11.11  & 0.1113  \T\\ \hline
4   & 0.0384 & 10.77 & 0.0749 & 0.0808 & 8.08 & 0   \T\\ \hline
5  & 0.0421 & 5.84 & 0.1194 & 0.1010  & 10.10 & 0.7227  \T\\ \hline
6 & 0.0510 & 11.02 & 0.0992 & 0.1010  & 10.10 & 0.0098 \T\\ \hline
7  & 0.0819 & 11.64 & 0.1305 & 0.1515   & 15.15 & 0.4107 \T\\ \hline
8   & 0.0465  & 0.93 & 0.1383 & 0.1010   & 10.10 & 1.5014 \T\\ \hline
9 & 0.0409 & 10.82 & 0.1132 & 0.0707  & 7.07 & 0   \T\\ \hline
10  & 0.1050  & 12.10 & 0.2918 & 0.0808  & 8.08 & 0   \T\\ \hline
\end{tabular}
\label{tab:Reserves_123}
\end{table*}

The characteristic functions shown in Table~\ref{tab:coalitions13} and \ref{tab:coalitions34} are used to calculate Shapley values using \eqref{eqn:shapley}, and the Shapley values are normalized based on \eqref{eqn:norm1} and \eqref{eqn:norm2}. The equivalent Shapley values are then computed using \eqref{eqn:eqShap} and finally the distribution factor ($DF$) of each DER is calculated using \eqref{eqn:DF}. For the case of the modified IEEE 13-node system, the distribution factors of DERs 1, 2, and 3, respectively, are 0.2729, 0.2185, and 0.5086. The allocated reserves of DERs (for $P_R = 100$ kW) are 34.09 kW, 8.28 kW, and 57.63 kW. These values are shown in Table~\ref{tab:Reserves_13}. Similarly, for the case of the modified IEEE 34-node system, the distribution factors of DERs 1, 2, 3, and 4, respectively, are 0.2949, 0.2141, 0.3499, and 0.1410. The allocated reserves of DERs (for $P_R = 100$ kW) are 28.85 kW, 16.42 kW, 20.50 kW, and 34.23 kW. These values are shown in Table~\ref{tab:Reserves_34}. Also, for the case of the modified IEEE 123-node system, the distribution factors of DERs and allocated reserves (for $P_R = 100$ kW) are shown in Table~\ref{tab:Reserves_123}.

The proposed approach is simulated on a PC with 64-bit Intel i5 core, 3.15 GHz processor, 8 GB RAM, and Windows OS. The execution time of the proposed approach for the IEEE 13-node, IEEE 34-node, and IEEE 123-node systems, respectively, are 0.46, 0.57, and 65.43 seconds. This indicates that the proposed approach is good enough for day-ahead (DA) as well as 15-minute ahead scheduling of reserves. 

\subsection{Comparison}
\begin{figure*}
    \centering    \includegraphics[scale=0.8]{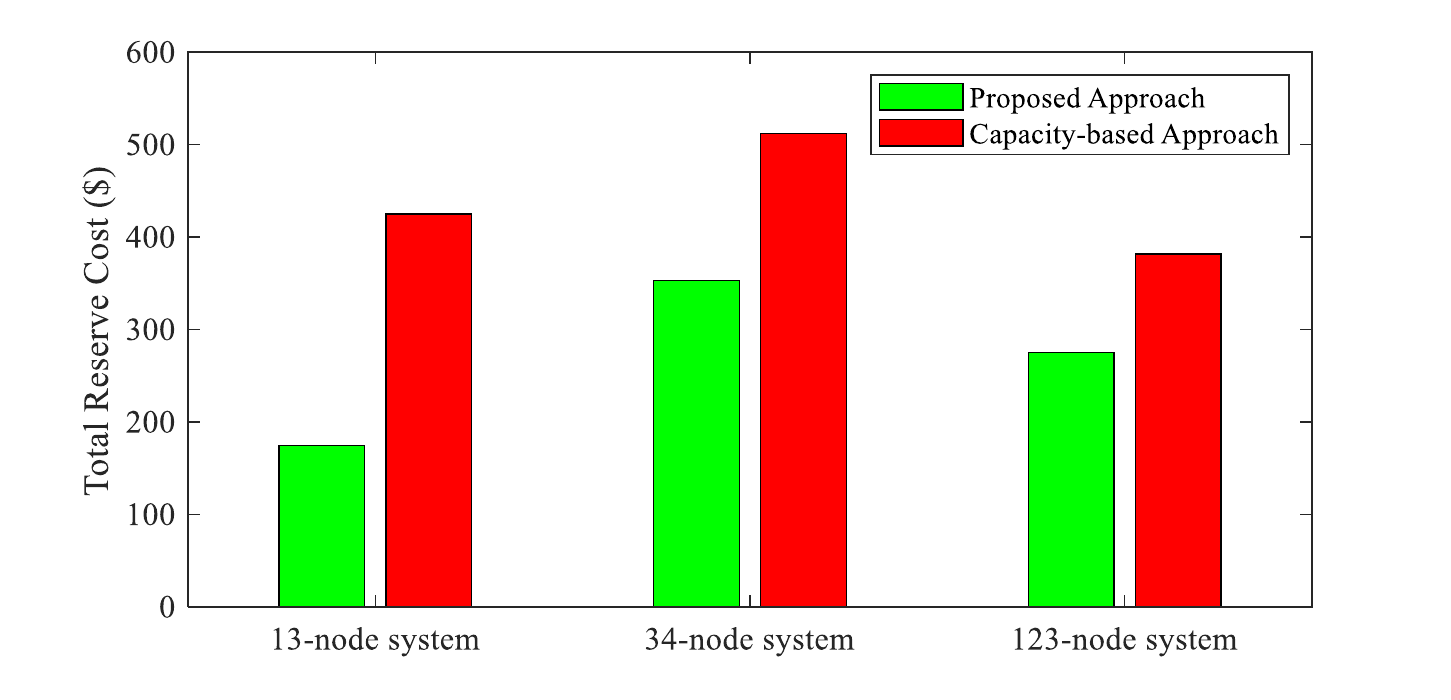}
    \caption{Bar graph showing the comparison of total reserve costs}
    \label{fig:bar}
\end{figure*}
The allocated reserves computed using the proposed approach are compared with that using a DER-capacity-based approach in terms of total reserve cost. In the DER-capacity-based approach, the distribution factors are computed based on sellable capacities of DERs. The distribution factors and the allocated reserves of DERs obtained using the DER-capacity-based approach for the modified IEEE 13-node, IEEE 34-node, and IEEE 123-node systems are, respectively, shown in Table~\ref{tab:Reserves_13}, Table~\ref{tab:Reserves_34}, and Table~\ref{tab:Reserves_123}. In case of the modified IEEE 13-node system, the total reserve costs obtained using the proposed approach and the DER-capacity-based approach, respectively, are \$174.57 and \$425.00. Similarly, in case of the modified IEEE 34-node system, the total reserve costs obtained using the proposed approach and the DER-capacity-based approach, respectively, are \$353.11 and \$511.82; and in case of the modified IEEE 123-node system, the total reserve costs obtained using the proposed approach and the DER-capacity-based approach, respectively, are \$274.94 and \$381.82. Figure~\ref{fig:bar} shows the comparison of total reserve costs obtained using the proposed approach and the DER-capacity-based approach for all test systems under consideration. The result shows that the total reserve cost can be lowered when the reserves are allocated using the proposed approach. The lowered cost of reserves is beneficial from the aggregator or system operator's perspective.

\begin{figure*}
    \centering
    \includegraphics[scale=0.95]{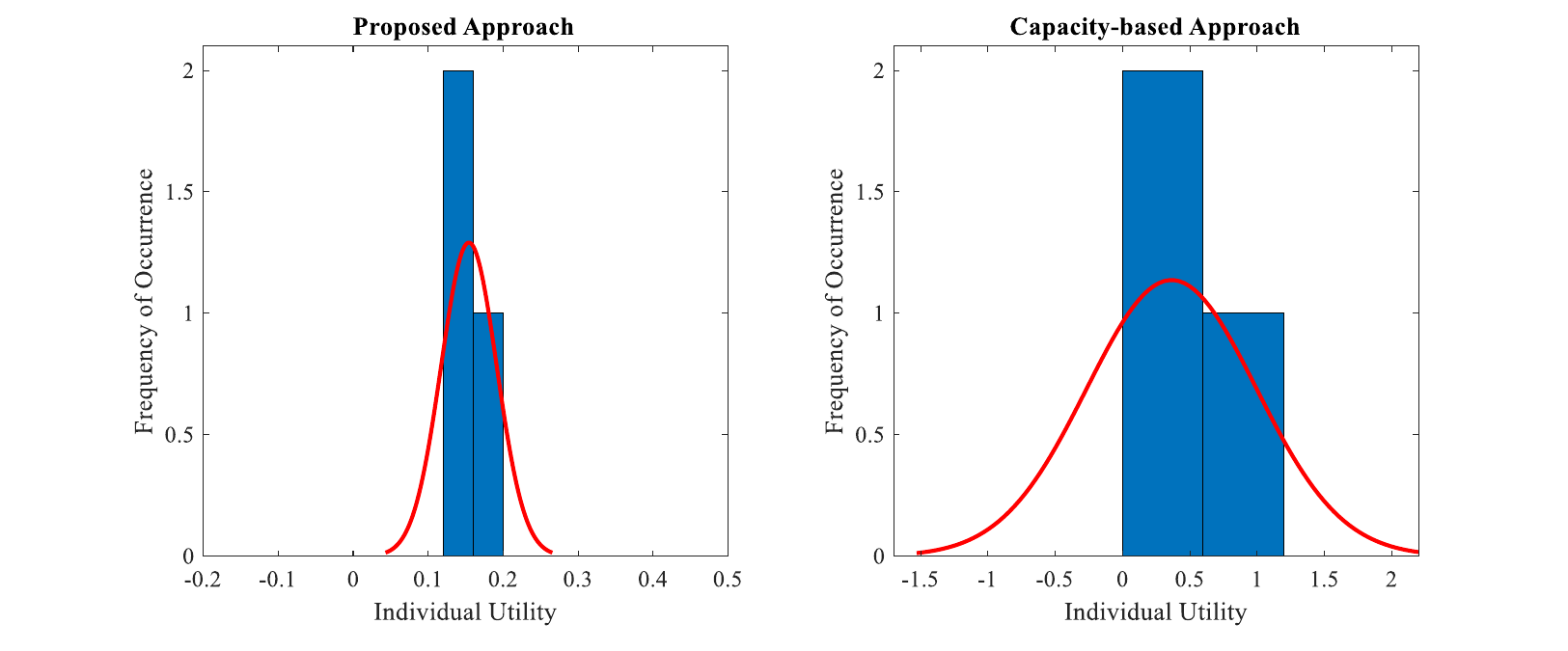}
    \caption{Histograms showing the distribution of individual utilities of DERs in case of the IEEE 13-node system}
    \label{fig:hist13}
\end{figure*}
\begin{figure*}
    \centering
    \includegraphics[scale=0.95]{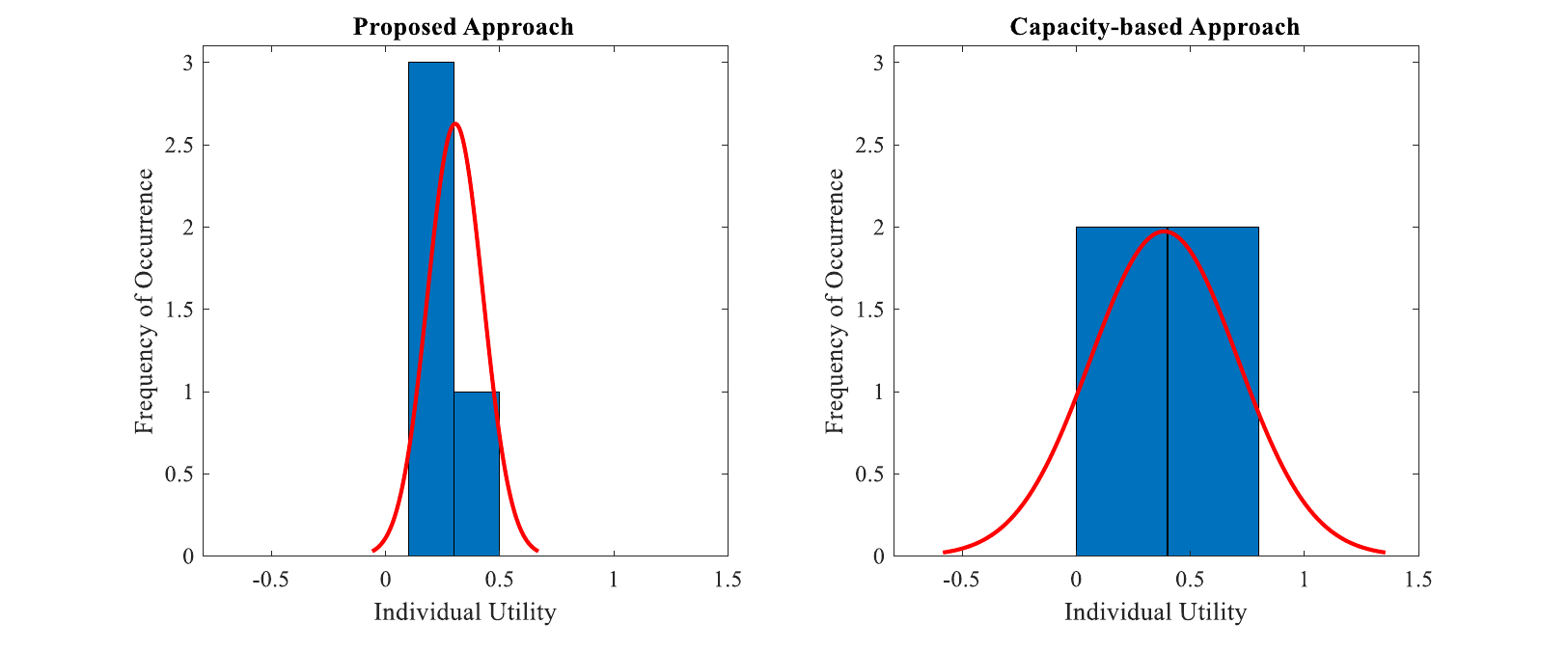}
    \caption{Histograms showing the distribution of individual utilities of DERs in case of the IEEE 34-node system}
    \label{fig:hist34}
\end{figure*}
\begin{figure*}
    \centering
    \includegraphics[scale=0.95]{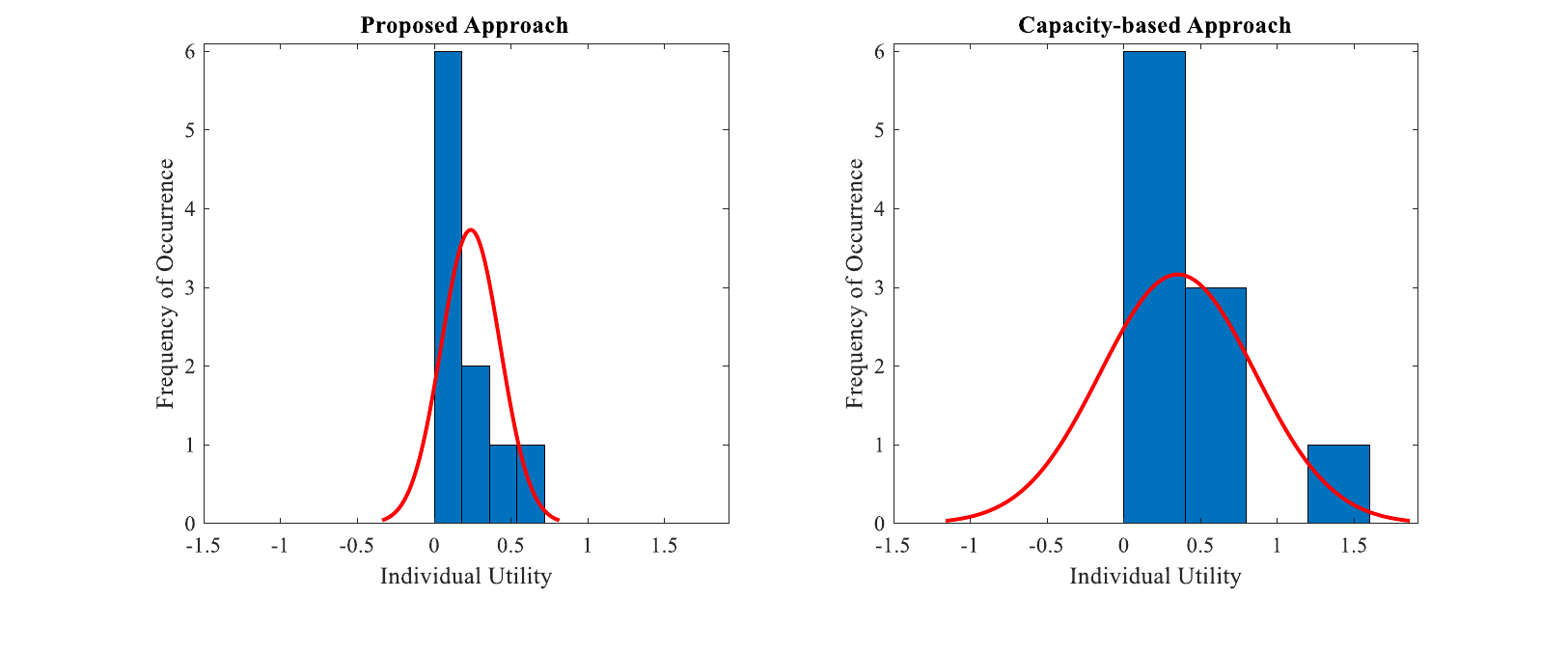}
    \caption{Histograms showing the distribution of individual utilities of DERs in case of the IEEE 123-node system}
    \label{fig:hist123}
\end{figure*}
To show the benefit of the proposed approach from the perspective of DER owners, the individual utility function is defined for the $i$\textsuperscript{th} DER as follows:
\begin{equation}
    u_i = 
    \frac{RBP_i \times \text{max}(0,AR_{i}-UCAR_i) \times PI_i}{P_{ci}} 
\end{equation}
where $RBP_i$ is the reserve bid price, $AR_i$ is the allocated reserve, $UCAR_i$ is the unpriced capacity available for reserve, $PI_i$ is the performance index, and $P_{ci}$ is the total sellable capacity of the $i$\textsuperscript{th} DER. 

Based on the values of allocated reserves, the individual utilities of DERs are calculated using both proposed and capacity-based approaches for all test systems under consideration, which are shown in Tables~\ref{tab:Reserves_13}, \ref{tab:Reserves_34}, and \ref{tab:Reserves_123}. Figures~\ref{fig:hist13}, \ref{fig:hist34}, and \ref{fig:hist123} show histograms of the individual utilities of DERs for IEEE 13-node, 34-node, and 123-node systems. From the figures, we can see that the individual utilities of DERs have less variance (or more closeness) for the proposed approach. The standard deviation of individual utilities of DERs for the proposed approach is 0.0371 and that for the capacity-based approach is 0.6316 in case of IEEE 13-node system. Similarly, the standard deviation of individual utilities of DERs for the proposed approach is 0.1214 and that for the capacity-based approach is 0.3232 in case of IEEE 34-node system; and the standard deviation of individual utilities of DERs for the proposed approach is 0.1924 and that for the capacity-based approach is 0.5040 in case of IEEE 123-node system. The more close (or less variant) individual utilities of DERs imply the similar level of satisfaction of DERs when the reserves are allocated based on the proposed approach. This is because of the use of Shapley values, which take marginal contribution of each DER into account while allocating the reserves.

\section{Conclusion}\label{conclusion}
In this paper, a coalitional game theoretic approach has been proposed to quantify potential participation of active distribution systems in tertiary frequency regulation. 
In the first stage, the two types of characteristic functions: worthiness index (WI) and power loss reduction (PLR) of each set of coalitions of DERs were computed. In the second stage, the equivalent Shapley values and distribution factors were computed for fair allocation of reserves among DERs and the determination of their active power set-points. In order to develop the proposed approach, the following variables/parameters of DERs have been taken into consideration: sellable capacity, market clearing capacity, performance index, and reserve bid price. The performance index was computed based on past historical data of committed power and supplied power of DERs. In order to demonstrate the effectiveness of the proposed approach, the case studies were performed on the modified IEEE 13-node, IEEE 34-node, and IEEE 123-node distribution test systems. The case study results exhibit the efficiency of the proposed approach for reserve allocation, which is beneficial to both system operator and individual DERs. 


\section*{Acknowledgement}
This article is based on work funded by the US Department of Energy's Office of Energy Efficiency and Renewable Energy (EERE) under Solar Energy Technologies Office Award Number DE-EE0009022.

\bibliographystyle{IEEEtran}
\bibliography{References.bib}

\begin{thebibliography}{10}
\providecommand{\url}[1]{#1}
\csname url@samestyle\endcsname
\providecommand{\newblock}{\relax}
\providecommand{\bibinfo}[2]{#2}
\providecommand{\BIBentrySTDinterwordspacing}{\spaceskip=0pt\relax}
\providecommand{\BIBentryALTinterwordstretchfactor}{4}
\providecommand{\BIBentryALTinterwordspacing}{\spaceskip=\fontdimen2\font plus
\BIBentryALTinterwordstretchfactor\fontdimen3\font minus
  \fontdimen4\font\relax}
\providecommand{\BIBforeignlanguage}[2]{{%
\expandafter\ifx\csname l@#1\endcsname\relax
\typeout{** WARNING: IEEEtran.bst: No hyphenation pattern has been}%
\typeout{** loaded for the language `#1'. Using the pattern for}%
\typeout{** the default language instead.}%
\else
\language=\csname l@#1\endcsname
\fi
#2}}
\providecommand{\BIBdecl}{\relax}
\BIBdecl

\bibitem{rapizza2020fast}
M.~R. Rapizza and S.~M. Canevese, ``Fast frequency regulation and synthetic
  inertia in a power system with high penetration of renewable energy sources:
  Optimal design of the required quantities,'' \emph{Sustainable Energy, Grids
  and Networks}, vol.~24, p. 100407, 2020.

\bibitem{zhou2019optimal}
Q.~Zhou, Z.~Tian, M.~Shahidehpour, X.~Liu, A.~Alabdulwahab, and A.~Abusorrah,
  ``Optimal consensus-based distributed control strategy for coordinated
  operation of networked microgrids,'' \emph{IEEE Transactions on Power
  Systems}, vol.~35, no.~3, pp. 2452--2462, 2019.

\bibitem{machowski2020power}
J.~Machowski, Z.~Lubosny, J.~W. Bialek, and J.~R. Bumby, \emph{Power system
  dynamics: stability and control}.\hskip 1em plus 0.5em minus 0.4em\relax John
  Wiley \& Sons, 2020.

\bibitem{perninge2017optimal}
M.~Perninge and R.~Eriksson, ``Optimal tertiary frequency control in power
  systems with market-based regulation,'' \emph{IFAC-PapersOnLine}, vol.~50,
  no.~1, pp. 4374--4381, 2017.

\bibitem{abbaspourtorbati2012towards}
F.~Abbaspourtorbati, M.~Scherer, A.~Ulbig, and G.~Andersson, ``Towards an
  optimal activation pattern of tertiary control reserves in the power system
  of \text{Switzerland},'' in \emph{2012 American Control Conference
  (ACC)}.\hskip 1em plus 0.5em minus 0.4em\relax IEEE, 2012, pp. 3629--3636.

\bibitem{malik2012decision}
O.~Mal{\'\i}k and P.~Havel, ``Decision support tool for optimal dispatch of
  tertiary control reserves,'' \emph{International Journal of Electrical Power
  \& Energy Systems}, vol.~42, no.~1, pp. 341--349, 2012.

\bibitem{delfino2002load}
B.~Delfino, F.~Fornari, and S.~Massucco, ``Load-frequency control and
  inadvertent interchange evaluation in restructured power systems,'' \emph{IEE
  Proceedings-Generation, Transmission and Distribution}, vol. 149, no.~5, pp.
  607--614, 2002.

\bibitem{donde2001simulation}
V.~Donde, M.~Pai, and I.~A. Hiskens, ``Simulation and optimization in an
  \text{AGC} system after deregulation,'' \emph{IEEE transactions on power
  systems}, vol.~16, no.~3, pp. 481--489, 2001.

\bibitem{bovera2021data}
F.~Bovera, G.~Rancilio, D.~Falabretti, and M.~Merlo, ``Data-driven evaluation
  of secondary-and tertiary-reserve needs with high renewables penetration: The
  \text{Italian} case,'' \emph{Energies}, vol.~14, no.~8, p. 2157, 2021.

\bibitem{querini2020cooperative}
P.~L. Querini, O.~Chiotti, and E.~Fern{\'a}dez, ``Cooperative energy management
  system for networked microgrids,'' \emph{Sustainable Energy, Grids and
  Networks}, vol.~23, p. 100371, 2020.

\bibitem{gautam2022coop_DER}
M.~Gautam, N.~Bhusal, and M.~Benidris, ``A cooperative game theory-based
  approach to sizing and siting of distributed energy resources,'' in
  \emph{2021 53rd North American Power Symposium (NAPS)}.\hskip 1em plus 0.5em
  minus 0.4em\relax IEEE, 2022, pp. 1--6.

\bibitem{gautam2022coop_SRF}
M.~Gautam, N.~Bhusal, M.~Benidris, and H.~Livani, ``A cooperative game
  theory-based secondary frequency regulation in distribution systems,'' in
  \emph{2021 53rd North American Power Symposium (NAPS)}.\hskip 1em plus 0.5em
  minus 0.4em\relax IEEE, 2022, pp. 1--6.

\bibitem{nazari2021economy}
M.~H. Nazari, M.~B. Sanjareh, A.~Khodadadi, M.~Torkashvand, and S.~H.
  Hosseinian, ``An economy-oriented dg-based scheme for reliability improvement
  and loss reduction of active distribution network based on game-theoretic
  sharing strategy,'' \emph{Sustainable Energy, Grids and Networks}, vol.~27,
  p. 100514, 2021.

\bibitem{rebours2007survey}
Y.~G. Rebours, D.~S. Kirschen, M.~Trotignon, and S.~Rossignol, ``A survey of
  frequency and voltage control ancillary services— \text{Part I}: Technical
  features,'' \emph{IEEE Transactions on power systems}, vol.~22, no.~1, pp.
  350--357, 2007.

\bibitem{gautam2020sensitivity}
M.~Gautam, N.~Bhusal, and M.~Benidris, ``A sensitivity-based approach to
  adaptive under-frequency load shedding,'' in \emph{2020 IEEE Texas Power and
  Energy Conference (TPEC)}.\hskip 1em plus 0.5em minus 0.4em\relax IEEE, 2020,
  pp. 1--5.

\bibitem{dorfler2015breaking}
F.~D{\"o}rfler, J.~W. Simpson-Porco, and F.~Bullo, ``Breaking the hierarchy:
  Distributed control and economic optimality in microgrids,'' \emph{IEEE
  Transactions on Control of Network Systems}, vol.~3, no.~3, pp. 241--253,
  2015.

\bibitem{gautam2021cooperative_PESGM}
M.~Gautam, N.~Bhusal, and M.~Benidris, ``A cooperative game theory-based
  approach to under-frequency load shedding control,'' in \emph{IEEE Power \&
  Energy Society General Meeting (PESGM)}.\hskip 1em plus 0.5em minus
  0.4em\relax IEEE, 2021, pp. 1--5.

\bibitem{wang2005operating}
J.~Wang, X.~Wang, and Y.~Wu, ``Operating reserve model in the power market,''
  \emph{IEEE Transactions on Power systems}, vol.~20, no.~1, pp. 223--229,
  2005.

\bibitem{gan2003energy}
D.~Gan and E.~Litvinov, ``Energy and reserve market designs with explicit
  consideration to lost opportunity costs,'' \emph{IEEE Transactions on Power
  Systems}, vol.~18, no.~1, pp. 53--59, 2003.

\bibitem{baringo2021offering}
L.~Baringo, M.~Freire, R.~Garc{\'\i}a-Bertrand, and M.~Rahimiyan, ``Offering
  strategy of a price-maker virtual power plant in energy and reserve
  markets,'' \emph{Sustainable Energy, Grids and Networks}, vol.~28, p. 100558,
  2021.

\bibitem{afshar2008method}
K.~Afshar, M.~Ehsan, M.~Fotuhi-Firuzabad, and N.~Amjady, ``A method for reserve
  clearing in disaggregated model considering lost opportunity cost,''
  \emph{Electric Power Systems Research}, vol.~78, no.~4, pp. 527--538, 2008.

\bibitem{pavic2019energy}
I.~Pavi{\'c}, Y.~Dvorkin, and H.~Pand{\v{z}}i{\'c}, ``Energy and reserve
  co-optimisation--reserve availability, lost opportunity and uplift
  compensation cost,'' \emph{IET Generation, Transmission \& Distribution},
  vol.~13, no.~2, pp. 229--237, 2019.

\bibitem{padmanabhan2019battery}
N.~Padmanabhan, M.~Ahmed, and K.~Bhattacharya, ``Battery energy storage systems
  in energy and reserve markets,'' \emph{IEEE Transactions on Power Systems},
  vol.~35, no.~1, pp. 215--226, 2019.

\bibitem{merten2020bidding}
M.~Merten, C.~Olk, I.~Schoeneberger, and D.~U. Sauer, ``Bidding strategy for
  battery storage systems in the secondary control reserve market,''
  \emph{Applied Energy}, vol. 268, p. 114951, 2020.

\bibitem{gautam2022cooperative}
M.~Gautam, N.~Bhusal, J.~Thapa, and M.~Benidris, ``A cooperative game
  theory-based approach to formulation of distributed slack buses,''
  \emph{Sustainable Energy, Grids and Networks}, vol.~32, p. 100890, 2022.

\bibitem{shapley1975competitive}
L.~S. Shapley and M.~Shubik, ``Competitive outcomes in the cores of market
  games,'' \emph{International Journal of Game Theory}, vol.~4, no.~4, pp.
  229--237, 1975.

\bibitem{curiel2013cooperative}
I.~Curiel, \emph{Cooperative game theory and applications: cooperative games
  arising from combinatorial optimization problems}.\hskip 1em plus 0.5em minus
  0.4em\relax Springer Science \& Business Media, 2013, vol.~16.

\bibitem{IEEEFEEDERS}
\BIBentryALTinterwordspacing
{Distribution System Analysis Subcommittee}, ``1992 test feeder cases,'' IEEE,
  PES, Tech. Rep., 1992. [Online]. Available:
  \url{http://sites.ieee.org/pestestfeeders/resources/}
\BIBentrySTDinterwordspacing

\end{thebibliography}

\end{document}